\documentclass[aps,prb,floatfix,superscriptaddress,showpacs,lengthcheck,hyperref,citeautoscript]{revtex4-1}
\usepackage{amsmath}
\usepackage{amssymb}
\usepackage{graphicx}
\usepackage{bm}
\usepackage{color}
\usepackage{MnSymbol}
\usepackage{verbatim}

\newcommand{\bea}{\begin{eqnarray}}
\newcommand{\eea}{\end{eqnarray}}
\newcommand{\be}{\begin{equation}}
\newcommand{\ee}{\end{equation}}

\newcommand{\tav}{t_\text{av}}
\newcommand{\trel}{t_{\text{rel}}}

\begin{document}
\title{Topological phase diagram of a three-terminal Josephson junction: From the conventional to the Majorana regime}

\author{Lucila {Peralta Gavensky}}
\affiliation{Centro At{\'{o}}mico Bariloche and Instituto Balseiro,
Comisi\'on Nacional de Energ\'{\i}a At\'omica (CNEA)- Universidad Nacional de Cuyo (UNCUYO), 8400 Bariloche, Argentina}
\affiliation{Instituto de Nanociencia y Nanotecnolog\'{i}a (INN-Bariloche), Consejo Nacional de Investigaciones Cient\'{\i}ficas y T\'ecnicas (CONICET), Argentina}

\author{Gonzalo Usaj}
\affiliation{Centro At{\'{o}}mico Bariloche and Instituto Balseiro,
Comisi\'on Nacional de Energ\'{\i}a At\'omica (CNEA)- Universidad Nacional de Cuyo (UNCUYO), 8400 Bariloche, Argentina}
\affiliation{Instituto de Nanociencia y Nanotecnolog\'{i}a (INN-Bariloche), Consejo Nacional de Investigaciones Cient\'{\i}ficas y T\'ecnicas (CONICET), Argentina}

\author{C. A. Balseiro}
\affiliation{Centro At{\'{o}}mico Bariloche and Instituto Balseiro,
Comisi\'on Nacional de Energ\'{\i}a At\'omica (CNEA)- Universidad Nacional de Cuyo (UNCUYO), 8400 Bariloche, Argentina}
\affiliation{Instituto de Nanociencia y Nanotecnolog\'{i}a (INN-Bariloche), Consejo Nacional de Investigaciones Cient\'{\i}ficas y T\'ecnicas (CONICET), Argentina}

\begin{abstract}
We study  the evolution of averaged transconductances in three-terminal Josephson junctions  when the superconducting leads are led throughout a topological phase transition from an $s$-wave to a $p$-wave (Majorana) phase by an in-plane magnetic field $B_x$. We provide a complete description of this transition as a function of $B_x$ and a magnetic flux $\Phi$ threading the junction. For that we use a spinful model within a formalism that allows to treat on equal footing the contribution to the transconductance from both the Andreev subgap levels and the continuum spectrum. We unveil a fractionalization in the quantization of the transconductance due to the presence of Majorana quasiparticles, reflecting the effective pumping of half a Cooper pair charge in the $p$-wave regime. 
\end{abstract}
\date{\today} 
\maketitle

\section{Introduction.}

The quest for finding novel topological properties in condensed matter systems has drawn increa\-sing attention both in theoretical and experimental research. Much of the progress in the field has relied on the idea of the emergence of topological features by en\-gi\-nee\-ring heterostructures of conventional materials~\cite{Alicea2012,Aguado2017,Sato2017,Ozawa2019}, dri\-ving them out of equilibrium~\cite{Oka2009,Kitagawa2010,Lindner2011} or by the direct design of topological phases with cold atoms systems~\cite{Bloch2012,Jotzu2014,Aidelsburger2014,Cooper2019}.
In this way, it is now possible to artificially mimic topological matter without the need of working with exotic materials per se.

Following this line of thought, multi-terminal Josephson junctions of at least four conventional superconductors coupled through a normal scattering region have been put forward as an exciting platform to engineer Weyl-topological phases~\citep{Riwar2016}. 
The existence of Weyl singularities in the spectrum of multiply connected networks can be traced back to the seminal work by Avron \textit{et al}~\cite{Avron1988}, who envisioned the possibility of non trivial quantization of averaged transconductances (ATCs) in these devices. Different Josephson junction (JJ) circuits were also suggested as practical realizations to experimentally probe topologically protected adiabatic transport in these artificial setups~\cite{Avron1989}.

More recently, it was shown that three terminal structures can also develop topological properties when piercing the central scattering region with an external magnetic flux~\cite{Meyer2017,Xie2017} or when driven with an external microwave field~\cite{PeraltaGavensky2018}. There have also been proposals to measure the Fubini-Study geometric tensor~\cite{Klees2018} and the Berry curvature~\cite{PeraltaGavensky2018,Klees2018} of a multi-terminal JJ using microwaves. Remarkably, all this theoretical activity has been fuelled by several stimulating experiments on multi-terminal superconducting devices~\cite{Strambini2016,Vischi2017,Draelos2019,Pankratova2018}.

In a multi-terminal JJ, ATCs can be experimentally probed by voltage biasing one of the superconducting reservoirs as a function of a control parameter and measuring the phase-averaged current in a different lead. For conventional $s$-wave pairing, this would lead to a transconductance quantization in units of $4e^2/h$~\cite{Eriksson2017}. 
 This outcome can be simply thought of as a non-trivial adiabatic charge pumping between leads, say $\nu$ and $\nu'$, a product of the existence of Weyl sources in the spectrum~\cite{Riwar2016,Avron1988}. For vanishing voltage $V$, the ATC results in $G_{\nu\nu'} = Q_P/(V\Delta T) = n_{\nu\nu'}2 e q^{\star}/h$,
where the pumping period $\Delta T$ is merely the Josephson period $h/2eV$ and the phase-averaged pumped charge $Q_P$ over this period is an integer multiple ($n_{\nu\nu'}\,\epsilon\,\mathbb{Z}$) of the effective charge of the Cooper pair $q^{\star} = 2e$. 

A natural follow up question is whether multi-terminal JJs with $p$-wave pairing that host Majorana fermions may also develop a quantized ATC. Interestingly enough, this would provide a robust measurement of the effective tunneling charge in the topological superconductors, which is expected to be $q^{\star} = e$  as in the case of the $4\pi$-periodic fractional Josephson effect ~\cite{Oreg2010, Lutchyn2010,Wiedenmann2016,Laroche2019}. 
A partial answer using the Kitaev spinless model was recently found in Ref.~[\onlinecite{Houzet2018}] for the case of a four-terminal device (see also~[\onlinecite{Xie2019}]).
Yet, a complete description of the transition from the $s$-wave to the $p$-wave regime within a spinful approach in the entire parameter space is still missing. This is a necessary condition to obtain a unified topological phase diagram that could be experimentally tested. Even more, the calculation of the topological invariant associated with the ATC has so far been mainly based on the Chern numbers of the Andreev bound states (ABS)~\cite{Riwar2016,Xie2017,Xie2019,Houzet2018,Deb2018} despite the fact that in many regions of the parameter space they cannot be disentangled from the continuum spectrum above the superconducting gap  (which might also contribute to the ATC~\cite{Meyer2017}).

\begin{figure}[t]
\includegraphics[width=.7\columnwidth]{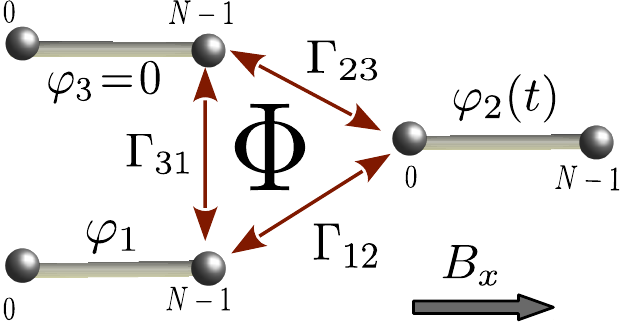}
\caption{Scheme of a Josephson tri-junction threaded by an external magnetic flux $\Phi$. The superconducting leads (with phases $\varphi_\nu$) have Rashba spin-orbit coupling and can be driven into a topological (Majorana) regime by an in plane magnetic field $B_x$.}
\label{fig1}
\end{figure}
In this work, we present the complete topological phase diagram of a three-terminal JJ obtained by a direct calculation of the adiabatic corrections to the Josephson current. We use the general approach developed in Ref.~[\onlinecite{PeraltaGavensky2018}] which includes the contributions from both the ABS and the continuum spectrum.
To describe the s-wave to p-wave transition, we consider a junction formed by spinful one-dimensional ($1D$) wires with Rashba spin-orbit coupling, proximity induced superconductivity and an in plane magnetic field $B_x$ (see Fig.~\ref{fig1}). This model, introduced in Refs.~[\onlinecite{Oreg2010}] and [\onlinecite{Lutchyn2010}], is known to have a topological phase transition that allows Majorana fermions to develop at the end of the wires for $B_x>B_c$, with $B_c$ a critical field. An external magnetic flux $\Phi$ threading the central region of the device  allows a fine tuning of the topological properties. 
For $B_x<B_c$ ($s$-wave phase) we recover the usual quantization ($4e^2/h$) whenever the system is gapped, while for $B_x>B_c$ ($p$-wave phase) we find a quantization in $2e^2/h$ that we identify with the development of effective spinless-like degrees of freedom, which make the measurement of a fractionalized pumped charge ($q^{\star} = e$) possible. All these results are summarized in Fig.~\ref{fig2}.

\section{Model Hamiltonian.}

The Hamiltonian of the Josephson tri-junction is given by $\hat{H}=\sum_{\nu=1}^{3}\hat{H}_{\nu}+\hat{H}_{ \mathcal{T}}$, where $\hat{H}_{\nu}$ is the Hamiltonian of the lead $\nu$, and $\hat{H}_{ \mathcal{T}}$ describes the Josephson tunneling. We use a spatially discretized version of the continuum model~\cite{Oreg2010, Lutchyn2010} with $N\rightarrow\infty$ lattice sites to describe the leads:
\begin{equation}
\hat{H}_{\nu}\!=\!\frac{1}{2}\!\sum_{j=0}^{N-1}\!\hat{\Psi}^{\nu\dagger}_{j}\mathcal{H}_{\nu}\hat{\Psi}^{\nu}_{j} + \!\frac{1}{2}\!\sum_{j=0}^{N-2}\!\Big[\hat{\Psi}^{\nu\dagger}_{j}T_{\nu}\hat{\Psi}^{\nu}_{j+1} + \hat{\Psi}^{\nu\dagger}_{j+1}T_{\nu}^{\dagger}\hat{\Psi}^{\nu}_{j}\Big],
\label{eqH}
\end{equation}
with  the four-component spinor in site $j$ defined as ${\hat{\Psi}^{\nu}_{j}}{} = (c_{j\uparrow}^{\nu},c_{j\downarrow}^{\nu},c_{j\downarrow}^{\nu\dagger},-c_{j\uparrow}^{\nu\dagger})^\mathrm{T}$ in the Nambu representation and 
\begin{eqnarray}
\notag
\mathcal{H}_{\nu}&=& (2t - \mu)\tau_z\otimes\sigma_0 + B_x \tau_0\otimes\sigma_x + \Delta \tau_x\otimes\sigma_0\,,\\
T_{\nu} &=& -t\tau_z\otimes\sigma_0 + i\alpha\tau_z\otimes\sigma_z.
\end{eqnarray}
where $t$ is the hopping matrix element, $\mu$ is the chemical potential, $\Delta$ is the superconducting order parameter, $\alpha$ is the spin-orbit coupling, and $B_x$ is the Zeeman field~\cite{note1}. The Pauli matrices $\tau_{a}$ ($\sigma_{a}$) and the identity $\tau_0$ ($\sigma_0$) act in particle-hole (spin) space. The junction's tunneling Hamiltonian $\hat{H}_{ \mathcal{T}}$ couples the leads only through their end site. To simplify the notation, we redefine $\hat{\Psi}_{N-1}^{1} \equiv \hat{\Psi}^{1}$, $\hat{\Psi}_{0}^{2} \equiv \hat{\Psi}^{2}$ and $\hat{\Psi}_{N-1}^{3} \equiv \hat{\Psi}^{3}$ and then write
\begin{eqnarray}
\label{tunnelling}
\hat{H}_{ \mathcal{T}} &=& \frac{1}{2}\sum_{\nu\leq\nu'}\Big[\hat{\Psi}^{\nu\dagger} \mathcal{T}_{\nu\nu'} \hat{\Psi}^{\nu'} + \hat{\Psi}^{\nu'\dagger} \mathcal{T}_{\nu\nu'}^{\dagger} \hat{\Psi}^{\nu} \Big]\,,\\
\notag
\mathcal{T}_{\nu\nu'} &=& \Big(\Gamma_{\nu\nu'}e^{i(\tau_z\otimes\sigma_0)[\frac{\varphi_\nu-\varphi_{\nu^{\prime}}}{2} - \epsilon_{\nu\nu'}\frac{\Phi}{3}]}+\delta_{\nu\nu'}U\Big)\tau_z\otimes\sigma_0,
\end{eqnarray}
with $\mathcal{T}_{\nu\nu'}^{\dagger}=\mathcal{T}_{\nu'\nu}$ and $\epsilon_{\nu\nu'}$ the totally antisymmetric tensor such that $\epsilon_{12} = \epsilon_{23} = \epsilon_{31} = 1$. The dimensionless parameter $\Phi$ describes a magnetic flux  threading the junction (in units of the flux quantum $\Phi_0 = h/e$) and $U$ is an on-site surface energy~\cite{note1}. 

\section{Topological averaged transconductance.}
\begin{figure*}[t]
\includegraphics[width=0.85\textwidth]{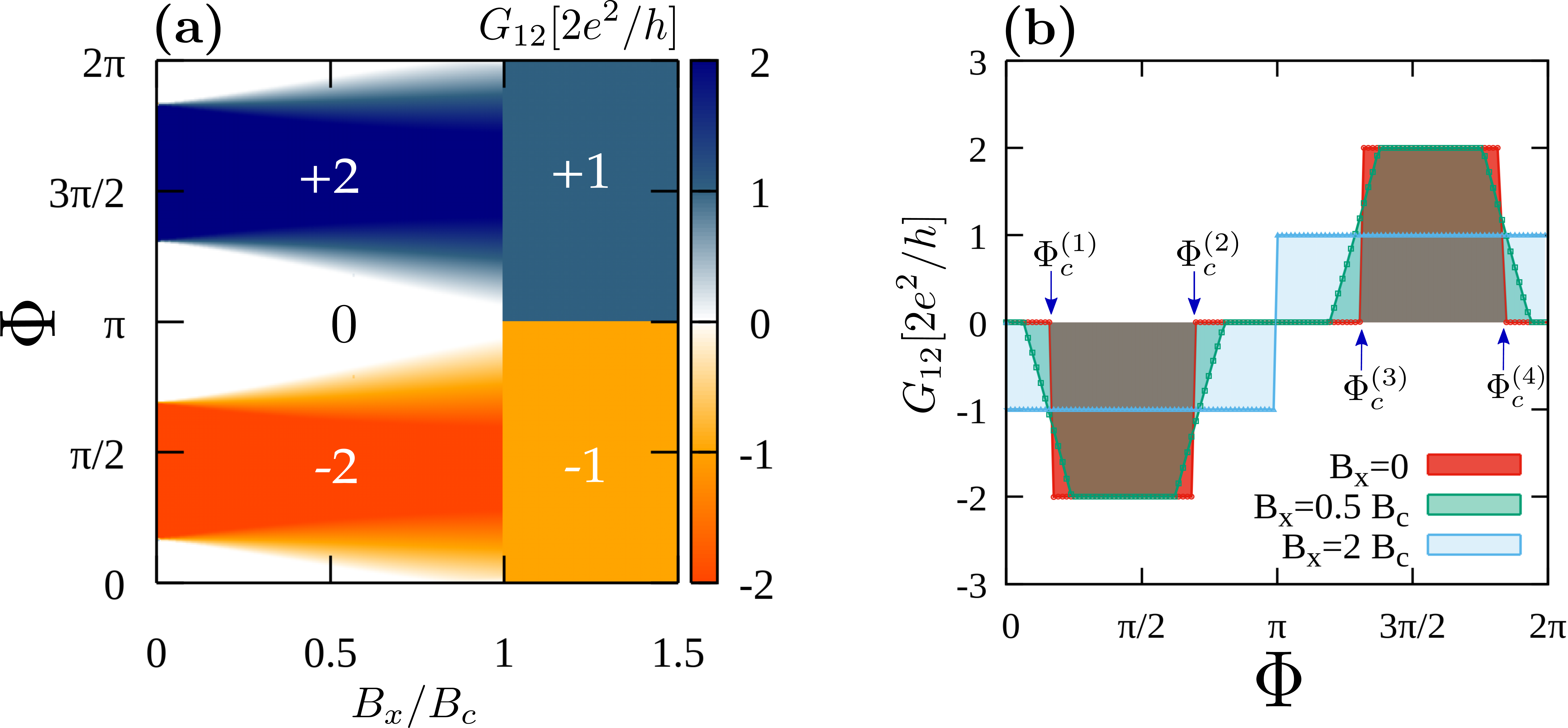}
\caption{(Color online) (a) Zero temperature average  transconductance $G_{12}$ as a function of the external flux $\Phi$  and the in-plane Zeeman field $B_x$ (in units of the critical field $B_c = \sqrt{\Delta^2+\mu^2}$). (b) $G_{12}$ as a function of the external flux $\Phi$ for different values of $B_x$. Notice that the effective pumped charge $q^{\star}$ changes from $2e$ to $e$ when $B_x$ becomes larger that $B_c$ (p-wave or Majorana phase).}
\label{fig2}
\end{figure*}
We are interested in computing the current $\langle \mathcal{J}_{\nu}\rangle$ flowing in lead $\nu$ when driving adiabatically the phase of a different lead $\nu'$ with a time dependence given by Josephson's relation $\dot{\varphi}_{\nu'} = 2e V_{\nu'}/\hbar$. The adiabatic correction to the current expectation value can be calculated using Wigner's representation of the two-time out-of-equilibrium Green's functions and performing a gradient expansion in the slow time scale (see Appendix A for details).
The ATC results in
\begin{equation}
\label{G}
G_{\nu\nu'} =\!\frac{1}{4\pi^2}\!\int_{0}^{2\pi}\!d\varphi_1\!\int_{0}^{2\pi}\!d\varphi_2\,\frac{\partial\langle \mathcal{J}_{\nu}\rangle}{\partial {V_{\nu'}}} = -\frac{2 e^2}{h}W_{\nu\nu'}
\end{equation}
where  
\begin{eqnarray}
\notag
W_{\nu\nu'}\!\!=\!\!\frac{1}{8\pi^2}\!\!\!\!\int\!\!\!\!d^2\varphi\!\!\int\!\!\!d\omega\!
\mathrm{Tr}\Big[ \mathcal{G}^{-1}\!\frac{\partial \mathcal{G}}{\partial{\varphi_{\nu}}}\!\cdot\mathcal{G}^{-1}\!\frac{\partial \mathcal{G}}{\partial{\varphi_{\nu'}}}\cdot\!\mathcal{G}^{-1}\!\frac{\partial \mathcal{G}}{\partial{\omega}}-\!\nu \leftrightarrow\nu'\!\Big],\\ 
\label{W}
\end{eqnarray}
is the winding number associated to the Green's function $\mathcal{G}$, which is topologically quantized whenever there is a gap at the Fermi energy. The trace is performed on both spin and particle-hole spaces.
In the above expressions, $\mathcal{G}(\omega,\{\varphi_\nu\})$ is the Feynman Green's function of the entire Hamiltonian in the total adiabatic limit---it follows the perturbation, $\{\varphi_\nu(t)\}$, instantaneously at time $t$---. This expression can be manipulated so as to be written in terms of the projected junction's Green function involving only the end sites of the nanowires ($\mathcal{G}_J$), see Appendix A. We compute $\mathcal{G}_J$ numerically by a simple decimation procedure  of the Hamiltonian defined in Eq.~(\ref{eqH}) (see Appendix B).
Eqs.~(\ref{G}) and (\ref{W}), originally derived in Ref.~[\onlinecite{PeraltaGavensky2018}], generalize the ones derived in Ref.~[\onlinecite{Riwar2016}] since they provide, on equal footing, the contribution to the ATC of both the ABS and the continuum states above the superconducting gap. In a more recent article, a similar expression in terms of the scattering matrix was obtained~\citep{Repin2019}.
There are two points worth mentioning: (i) If both spin projections are degenerate, then the trace over these variables would simply result in a factor $2$, which means a transconductance quantization in units of $4e^2/h$, and (ii) the Green's function formalism implicitly assumes a thermodynamic average (taken here at zero temperature), which means that if there is a band crossing of the ABS at zero energy, the  transconductance will not necessarily be quantized.

\subsection{Topological phase diagram.}

Figure \ref{fig2} summarizes the main result of this work: (a) a color map of the thermal transconductance as a function of the external flux $\Phi$ and the in-plane Zeeman field $B_x$; (b) representative cuts of this color map for fixed $B_x$. 
For $B_x = 0$ both spin projections are degenerate and four zero energy Weyl points develop as a function of the external flux at $\Phi_c^{(i)}$ for $i=1,2,3,4$, marked with arrows in Fig.~\ref{fig2}(b), making the ATC jump at these singular points between quantized values of $\pm 4e^2/h$ and zero. 
As $B_x$ increases, the quantization does not persist at all flux values, which can only be due to band overlaps developing in the spectrum at zero energy, as we shall show below.
At this point, we would like to emphasize that in this entire region ($B_x<B_c$) the independent contributions to the ATC of both the ABS and the continuum spectrum are non trivial and that our method adds them up to give the correct value. In some cases, it is possible to separate them by resorting to the so-called topological Hamiltonian, defined as $H_\mathrm{top}^{J}(\varphi_1,\varphi_2)\equiv-\mathcal{G}_J^{-1}(\omega=0)$, which has been proven~\citep{Wang2012,Wang2013} to have the same topological properties as the system described by the entire $\mathcal{G}_J(\omega)$. Indeed, the Chern numbers calculated from the spectrum of $H_\mathrm{top}^J$ do agree (see Appendix C) with the results presented in Fig.~\ref{fig2} and, for the particular case of $B_x=0$, with those obtained in Ref.~[\onlinecite{Meyer2017}].

The most striking result shown in Fig.~\ref{fig2} is manifested for $B_x > B_c$, where the ATC only takes quantized values in $2e^2/h$. For these fields, Majorana degrees of freedom are expected to emerge in the system, with an effectively spinless superconductivity that causes the effective charge $q^{\star}$ to be that of the single electron.

\subsection{Junction's spectral density.}
\begin{figure}[b]
\includegraphics[width=0.9\columnwidth]{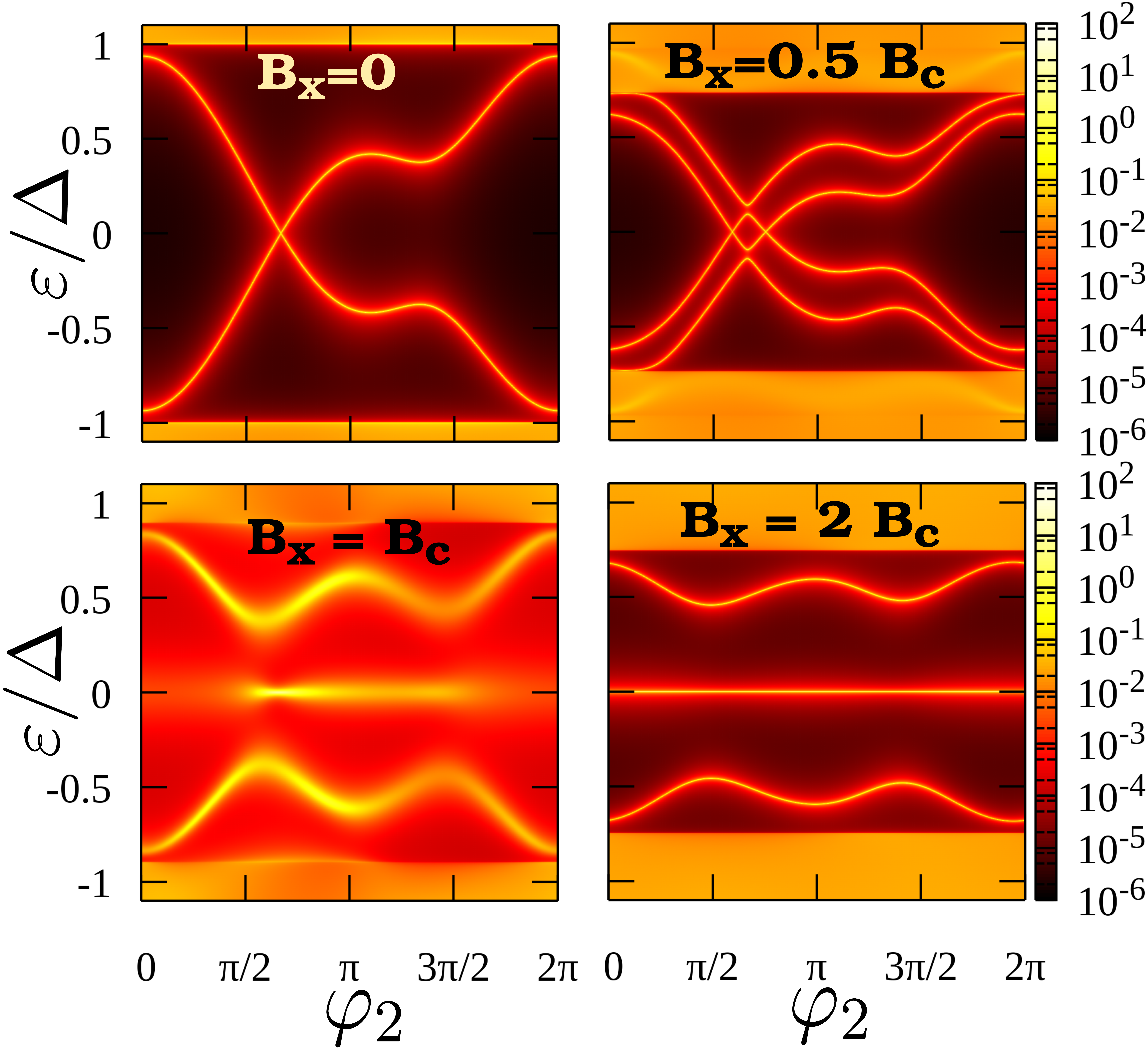}
\caption{(Color online) Andreev subgap spectrum along the path $\varphi_2 = -\varphi_1$ as a function of the Zeeman field $B_x$ for $\mu=3.2\Delta$ and spin orbit coupling $\alpha=50\Delta$. The central magnetic flux was taken to be $\Phi_c^{(1)}=0.53$.}
\label{fig3}
\end{figure}

In order to understand these results, we show in Fig. \ref{fig3} the spectral density of the tri-junction $\mathcal{A}(\omega)=-1/\pi\, \mathrm{Im}[\mathrm{Tr}(\mathcal{G}^r_J(\omega))]$ along the path $\varphi_2 = -\varphi_1$ for different values of  $B_x$, where $\mathcal{G}^r_J(\omega)$ stands for the retarded junction Green's function. The external magnetic flux was taken to be $\Phi_c^{(1)} = 0.53$, which corresponds to the first critical flux where a Weyl point appears in the ABS spectrum at zero magnetic field. As $B_x$ increases, the spin degeneracy is lifted and the ABSs overlap--as discussed in Ref.~[\onlinecite{Yokoyama2015}]--causing the ATC to be not quantized. At the critical field $B_c = \sqrt{\Delta^2 + \mu^2}$ the superconducting gap closes, and for $B_x > B_c$ a zero energy flat band emerges. The existence of this dispersionless state can be traced, as discussed below, to the fact that in the $p$-wave regime, Majorana modes $\hat{\gamma}_\nu$ develop at the end of each lead $\nu$. Only two effectively spinless dispersive ABSs coexist inside the gap with the Majorana band. Notice that in this three-terminal setup, no further requirement beyond thermal occupation is needed in order to observe the $2e^2/h$ quantization since for any flux aside from the critical ones the two ABSs never touch in the Majorana phase.

\begin{figure}[t]
\includegraphics[width=0.9\columnwidth]{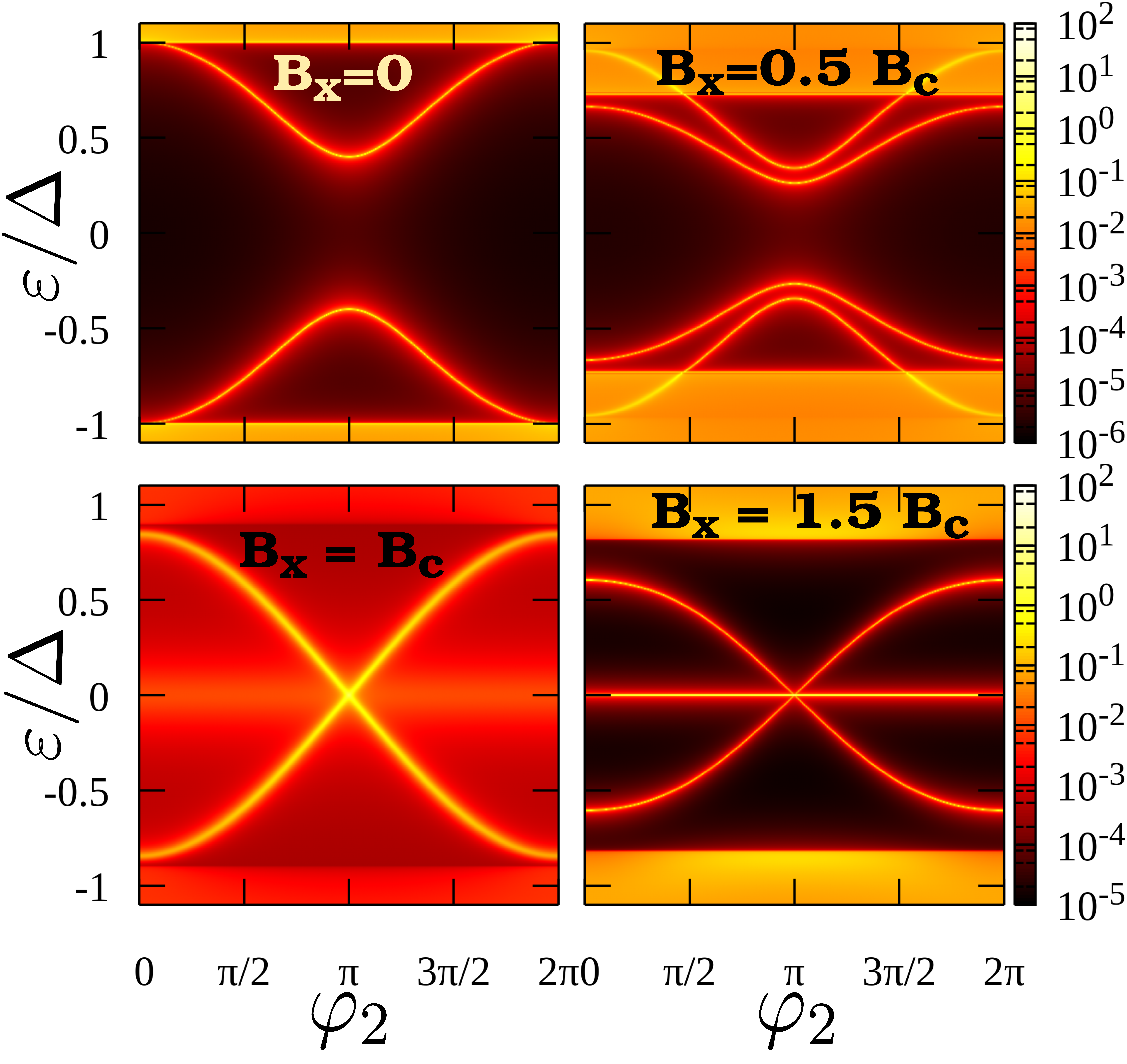}
\caption{Andreev subgap spectrum along $\varphi_2$ for $\varphi_1 = 2\pi$ as a function of the Zeeman field $B_x$ for $\mu=3.2\Delta$ and spin orbit coupling $\alpha=50\Delta$. The central magnetic flux was taken to be $\Phi=0$.}
\label{fig4}
\end{figure}

In Fig. \ref{fig4} we show the spectral density along a different path ($\varphi_1=2\pi$) as a function of the Zeeman field for no external flux ($\Phi = 0$). Here, it becomes clear that the jump in the ATC from $-2e^2/h$ to $2e^2/h$ for $B_x>B_c$ can be traced back to the emergence of a pseudo-spin $1$ Dirac-Weyl point, as it is apparent in the figure for $B_x = 1.5\,B_c$. In this condition, when the external magnetic flux is turned on, it breaks the degeneracy between the two ABS, making them detach from the zero energy (Majorana) solution. 

We notice that there are situations in which the ABSs are not detached from the continuum  but they merge into it in the form of resonances (see Figs.~\ref{fig3} and \ref{fig4})---this is so even in the Majorana regime (not shown).  
These mergings do not change the Green's function winding number since when two states become degenerate the Chern number of each one may change but their sum remains constant~\cite{Avron1983}. Hence, all changes in the quantized ATC presented in the topological phase diagram of Fig.~\ref{fig2} are due to the occurrence of crossings or Weyl points in the Andreev spectrum at zero energy.

In the $p$-wave phase ($B_x>B_c$), the topological Hamiltonian $H_\mathrm{top}$ is not well defined due to the presence of the Majorana bound state. However, for $B_x \gg B_c$, i.e. deep into the Kitaev regime, a simple low-energy model helps to describe and analyze the results. In this case, the tunneling Hamiltonian takes the form  
\begin{equation}
H_T^{\mathrm{eff}}(\varphi_1,\varphi_2) = i t_{12}\hat{\gamma}_{1}\hat{\gamma}_{2} + i t_{23}\hat{\gamma}_{2}\hat{\gamma}_{3} + i t_{31}\hat{\gamma}_{3}\hat{\gamma}_{1} ,
\end{equation}
with $t_{12} = \tilde{\Gamma}_{12}\cos((\varphi_1-\varphi_2)/2-\Phi/3)$, $t_{23}=-\tilde{\Gamma}_{23}\cos(\varphi_2/2-\Phi/3)$, $t_{31}=-\tilde{\Gamma}_{13}\sin(\varphi_1/2+\Phi/3)$ and $\tilde{\Gamma}_{ij}=\Gamma_{ij}/2$. It  possesses two particle-hole symmetric eigenvectors $|\psi_{\pm}\rangle$ with energies $E_{\pm} = \pm \sqrt{t_{12}^2 + t_{13}^2 + t_{23}^2}$, which correspond to a finite energy fermion (two Majorana degrees of freedom are paired up), and one zero-energy solution (the flat band) that is merely an ``unpaired" Majorana mode living in the junction.
Within this approximation it is much easier to show that all the winding comes from the finite energy fermionic band whose Chern number 
\begin{equation}
C_{-}\!=\!\frac{1}{8\pi}\!\int_{0}^{4\pi}\!d\varphi_1\!\int_{0}^{4\pi}\!\!d\varphi_2 \Big[i\langle\partial_{\varphi_1}\psi_{-}|\partial_{\varphi_2}\psi_{-}\rangle - (\varphi_1\! \leftrightarrow\!\varphi_2)\Big]
\end{equation}
changes from $-1$ to $1$ at $\Phi =0$ and $\Phi =\pi$. In this regime, the continuum does not contribute.
To further check these results, we have also made use of an analytic expression for the boundary Green's function for a Kitaev chain in the wide band limit~\cite{Zazunov2016}
to calculate the winding number $W_{\nu\nu'}$. Interestingly, we do recover the same results as in the spinful model, even though in this case, the continuum has a smoother dependence in energy than in the complete model. 

\section{Conclusions.}

By expressing the average transconductance of a multi-terminal JJ in terms of the Green's function of the junction, we have calculated the topological phase diagram of a three-terminal junction that completely describes the transition from the $s$-wave to the $p$-wave regime within a simple spinful model.  In the former case we found that transconductance is quantized in units of $4e^2/h$, whenever the ABS spectrum is gapped, while in the latter it is in units of $2e^ 2/h$. This is  consistent with a change of the pumped charge  over a Josephson period from $2e$ to $e$, a distinctive signature of the fractional Josephson effect. Notably, in our odd-terminal setup this occurs under thermal occupation conditions---this is not the case in $4$-terminal JJ, see~[\onlinecite{Houzet2018}].

As mentioned above, the Green's function formalism used to calculate the winding number turns out to be essential to properly account for the contribution of both the Andreev subgap levels and the continuum states, which are, in general, unavoidably mixed. In addition, this method  possesses several important advantages. On the one hand, it can in principle incorporate the effect of interactions, provided a good approximation for the single-particle Green's function is obtained. It can also properly handle the presence of a uncoupled Majorana state always pinned at zero energy.
On the the other hand, the method could easily incorporate microscopic details of the junction through more realistic calculations of the junction's Green's functions that might be relevant for a proper comparison with experimental data.

\begin{acknowledgments}
We acknowledge financial support from ANPCyT (grants PICTs 2013-1045 and 2016-0791), from CONICET (grant PIP 11220150100506) and from SeCyT-UNCuyo (grant 06/C526).
\end{acknowledgments}

\appendix
\section{Adiabatic expansion of the Josephson's current.}

We present here, for the sake of completeness, the expression derived in Ref.~[\onlinecite{PeraltaGavensky2018}] to calculate the topological transconductance of a multi-terminal Josephson junction. In general, the adiabatic expansion of the Josephson current can be done directly with the Green's function of the junction's Hamiltonian. We begin with Dyson's equation of motion of the Green's function of the entire (non-interacting) Hamiltonian $H(t)$ describing the junction and the leads written in the Bogoliubov-de Gennes (BdG) representation of particle-hole and spin spaces used in the main text
\begin{equation}
[i\partial_t - H(t)]\hat{\mathcal{G}}(t,t') = \delta(t-t'),
\label{eqDyson}
\end{equation}
where the time dependence appears only on the tunneling Hamiltonian throughout the superconducting phases. The approach is based on the separation of slow and fast timescales in Dyson's equation by taking advantage of Wigner's representation. Under this formalism, two time variables, the relative time $t_{\text{rel}}=t-t'$ and the average one $t_{\text{av}}=\frac{t+t'}{2}$, are handled. Time derivatives of the former are used as a small parameter, making feasible a perturbation scheme. Wigner's representation is given by
\begin{eqnarray}
\notag
\hat{\mathcal{G}}(t,t') &=& \int_{-\infty}^{\infty}\frac{d\omega}{2\pi}e^{-i\omega (t-t')}\widetilde{\mathcal{G}}(\omega,t_{\text{av}}),\\
\widetilde{\mathcal{G}}(\omega,t_{\text{av}}) &=& \int_{-\infty}^{\infty}dt_{\text{rel}}e^{i\omega t_{\text{rel}}}\hat{\mathcal{G}}(t,t').
\end{eqnarray}
Taking into account that $\partial_t = \frac{1}{2}\partial_{t_{\text{av}}} + \partial_{t_{\text{rel}}}$, Eq. (\ref{eqDyson}) can be written to all orders as~[\onlinecite{Kershaw2017}]
\begin{equation}
[\omega + \frac{i}{2}\partial_{\tav} - e^{\frac{1}{2i}\partial_{\omega}^{\mathcal{G}}\partial_{\tav}^{H}}H(\tav)]\widetilde{\mathcal{G}}(\omega,\tav) =
\mathcal{I},
\label{eqDyson-Wigner}
\end{equation}
where we applied the following manipulation
\begin{eqnarray}
\notag
\int_{-\infty}^{\infty}e^{i\omega\trel}H(\tav + \trel/2)\hat{\mathcal{G}}\big(\tav+\frac{\trel}{2}, \tav - \frac{\trel}{2}\big)d\trel=\\
\notag
\int_{-\infty}^{\infty}e^{i\omega\trel}e^{\frac{\trel}{2}\partial_{\tav}^{H}}H(\tav)\hat{\mathcal{G}}\big(\tav+\frac{\trel}{2}, \tav - \frac{\trel}{2}\big)d\trel=\\
\notag
\int_{-\infty}^{\infty}e^{\frac{1}{2 i}\partial_{\omega}\partial_{\tav}^{H}}e^{i\omega\trel} H(\tav)\hat{\mathcal{G}}\big(\tav+\frac{\trel}{2}, \tav - \frac{\trel}{2}\big)d\trel=\\
\notag
=e^{\frac{1}{2i}\partial_{\omega}^{\mathcal{G}}\partial_{\tav}^{H}}H(\tav)\widetilde{\mathcal{G}}(\omega,\tav).\\
\end{eqnarray}
We are interested in a first order approximation, where the Green's function is perturbed up to first derivatives of the average time. In that case, Eq. (\ref{eqDyson-Wigner}) is given by
\begin{equation}
\Big(\omega +\frac{i}{2}\partial_{\tav} - H(\tav) -\frac{1}{2i}\frac{\partial H(\tav)}{\partial \tav}\partial_{\omega}\Big)\widetilde{\mathcal{G}}(\omega,\tav)=\mathcal{I}.
\end{equation}
Introducing a series expansion for the Green's function $\widetilde{\mathcal{G}}(\omega,t_{\text{av}}) = \sum_n\widetilde{\mathcal{G}}^{(n)}$, we can obtain the next system of equations
\begin{eqnarray}
\label{eq-a}
\Big(\omega - H\Big)\widetilde{\mathcal{G}}^{(0)} &=& \mathcal{I},\\
\label{eq-b}
\Big(\omega - H\Big)\widetilde{\mathcal{G}}^{(1)} &=& \frac{1}{2i}\partial_{\tav}H\partial_{\omega}\widetilde{\mathcal{G}}^{(0)} + \frac{1}{2i}\partial_{\tav}\widetilde{\mathcal{G}}^{(0)},
\end{eqnarray}
where the arguments $(\omega,\tav)$ are omitted for the sake of brevity. Solving Eq. (\ref{eq-a}) and replacing into Eq. (\ref{eq-b}) we get
\begin{eqnarray}
\label{eq-c}
\widetilde{\mathcal{G}}^{(0)} &=& \Big(\omega - H\Big)^{-1},\\
\label{eq-d}
\widetilde{\mathcal{G}}^{(1)} &=& \frac{1}{2i}\widetilde{\mathcal{G}}^{(0)}\Big[\widetilde{\mathcal{G}}^{(0)},  \partial_{\tav}H\widetilde{\mathcal{G}}^{(0)}\Big],
\end{eqnarray}
where $\Big[\cdot,\cdot\Big]$ is the commutator. We've also used the properties of the derivatives of inverse functions $\partial_{\lambda}A^{-1} = -A^{-1}\partial_{\lambda}A A^{-1}$, meaning in this case that
\begin{eqnarray}
\partial_\omega\widetilde{\mathcal{G}}^{(0)} &=& -\widetilde{\mathcal{G}}^{(0)}\widetilde{\mathcal{G}}^{(0)},\\
\partial_{\tav}\widetilde{\mathcal{G}}^{(0)} &=& \widetilde{\mathcal{G}}^{(0)}\partial_{\tav}H\widetilde{\mathcal{G}}^{(0)}.
\end{eqnarray}
The mean value of a single particle operator written in the BdG basis $\hat{\mathcal{O}}(t) = \frac{1}{2}\sum_{\alpha\beta}\hat{\Psi}^{\dagger}_{\beta}(t)\mathcal{O}_{\beta\alpha}\hat{\Psi}_{\alpha}(t)$ is given by
\begin{eqnarray}
\notag
\langle\phi_{0}|\hat{\mathcal{O}}(t)|\phi_{0}\rangle &=& \frac{1}{2}\lim_{t'-t\to\epsilon^{+}}\sum_{\alpha\beta}\mathcal{O}_{\beta\alpha}\langle\phi_{0}|\hat{\Psi}^{\dagger}_{\beta}(t')\hat{\Psi}_{\alpha}(t)|\phi_{0}\rangle\\
\notag
&=& -\frac{i}{2}\lim_{t'-t\to\epsilon^{+}}\sum_{\alpha\beta}\mathcal{O}_{\beta\alpha}\mathcal{G}_{\alpha\beta}(t,t')\\
&=& -\frac{i}{2}\lim_{t'-t\to\epsilon^{+}}\text{Tr}\Big[\mathcal{O}\hat{\mathcal{G}}(t,t')\Big]
\label{eqmean-value}
\end{eqnarray}
By writing Eq. (\ref{eqmean-value}) in Wigner's represention and expanding the Green's function up to first order we obtain
\begin{eqnarray}
\langle\mathcal{O}(t)\rangle\!&\simeq&\!\frac{1}{2}\lim_{t'-t\to\epsilon^{+}}\Big\{-i\int_{-\infty}^{\infty}\frac{d\omega}{2\pi} \text{Tr}\Big[\mathcal{O}\widetilde{\mathcal{G}}^{(0)}\Big]\\
\notag
\!&+&\!\int_{-\infty}^{\infty}\!\frac{d\omega}{4\pi}\!\text{Tr}\Big(\!\frac{\partial H}{\partial \tav}\widetilde{\mathcal{G}}^{(0)}\mathcal{O}\frac{\partial \widetilde{\mathcal{G}}^{(0)}}{\partial \omega} -\!\mathcal{O}\widetilde{\mathcal{G}}^{(0)}\frac{\partial H}{\partial \tav}\frac{\partial \widetilde{\mathcal{G}}^{(0)}}{\partial \omega}\!\Big)\!\Big\}.
\end{eqnarray}
By specifying $\mathcal{O}$ as the current operator $\mathcal{J}_{\nu} = 2e\partial_{\varphi_{\nu}}H$ and using that $\lim_{t'-t\to\epsilon^{+}}\partial_{\tav}H = \sum_{\nu'}\partial_{\varphi_{\nu'}}H\dot{\varphi}_{\nu'}(t)$, we get the first order correction to the current expectation value given by
\begin{widetext}
\begin{eqnarray}
\notag
\langle\mathcal{J}^{(1)}_{\nu}(t)\rangle &=& e\sum_{\nu'}\int_{-\infty}^{\infty}\frac{d\omega}{4\pi}\text{Tr}\Big[\partial_{\varphi_{\nu'}}H\cdot\widetilde{\mathcal{G}}^{(0)}\partial_{\varphi_{\nu}}H\cdot\partial_{\omega}\widetilde{\mathcal{G}}^{(0)}-\nu' \leftrightarrow\nu\Big]\dot{\varphi}_{\nu'}(t)\\
&=& -\frac{2e^2}{\hbar}\sum_{\nu'}\int_{-\infty}^{\infty}\frac{d\omega}{4\pi}\text{Tr}\Big[\widetilde{\mathcal{G}}^{(0)-1}\partial_{\varphi_{\nu}}\widetilde{\mathcal{G}}^{(0)}\cdot\widetilde{\mathcal{G}}^{(0)-1}\partial_{\varphi_{\nu'}}\mathcal{G}^{(0)}\cdot\widetilde{\mathcal{G}}^{(0)-1}\partial_{\omega}\widetilde{\mathcal{G}}^{(0)}-\nu \leftrightarrow\nu'\Big]V_{\nu'}.
\end{eqnarray}
\end{widetext}
The averaged transconductance then simply results in
\begin{equation}
G_{\nu\nu'} =\!\frac{1}{4\pi^2}\!\int_{0}^{2\pi}\!d\varphi_1\!\int_{0}^{2\pi}\!d\varphi_2\,\frac{\partial\langle\mathcal{J}^{(1)}_{\nu}\rangle}{\partial {V_{\nu'}}},
\end{equation}
which leads to Eq.~(\ref{G}) used to obtain our main results. We then recover an expression of this adiabatic transport coefficient in terms of the Green's function winding number. 
By noticing that $\widetilde{\mathcal{G}}^{(0)}(\omega,t_\mathrm{av})$ depends on the average time $t_\mathrm{av}$ only throughout the superconducting phases we can define $\mathcal{G}(\omega,\{\varphi_\nu\})=\widetilde{\mathcal{G}}^{(0)}(\omega,\{\varphi_\nu(t_\mathrm{av})\})$.
Furthermore, since the tunneling Hamiltonian acts locally on the end sites of each lead, this expression can be simplified so as to be written only in terms of the projected junction Green's function $\mathcal{G}_J$.
The averaged transconductance can then be written as
\begin{eqnarray}
G_{\nu\nu'} &=&-\frac{2e^2}{h}\frac{1}{8\pi^2}\!\int_{0}^{2\pi}\!d^2\varphi\int d\omega\\
\notag
&\times&\text{Tr}_J\Big[\partial_{\varphi_{\nu'}}\mathcal{T}\cdot\mathcal{G}_J\cdot\partial_{\varphi_{\nu}}\mathcal{T}\cdot\partial_{\omega}\mathcal{G}_J-\nu' \leftrightarrow\nu\Big],
\end{eqnarray}
where the phase derivatives have been explicitly written in terms of the tunneling Hamiltonian $\mathcal{T}$ as defined in Eq.~(\ref{tunnelling}), and the trace $\text{Tr}_J$ acts locally on the junction's states. We emphasize that the result does not depend explicitly on the slow time-scale any more, since the instantaneous superconducting phases are integrated out. 

\section{Boundary Green's functions of the uncoupled leads by decimation.}

The spinful boundary Green's function of each one of the leads is obtained by a numerical decimation procedure applied to the Hamiltonian
\begin{equation}
\hat{H}\!=\!\frac{1}{2}\!\sum_{j=0}^{N-1}\!\hat{\Psi}^{\dagger}_{j}\mathcal{H}\hat{\Psi}_{j} + \!\frac{1}{2}\!\sum_{j=0}^{N-2}\!\Big[\hat{\Psi}^{\dagger}_{j}T\hat{\Psi}_{j+1} + \hat{\Psi}^{\dagger}_{j+1}T^{\dagger}\hat{\Psi}^{}_{j}\Big],
\label{eq1}
\end{equation}
where the four component spinor in site $j$ is defined as ${\hat{\Psi}_{j}}{} = (c_{j\uparrow},c_{j\downarrow},c_{j\downarrow}^{\dagger},-c_{j\uparrow}^{\dagger})^\mathrm{T}$ in the Nambu representation and 
\begin{eqnarray}
\notag
\mathcal{H}&=& (2t - \mu)\tau_z\otimes\sigma_0 + B_x \tau_0\otimes\sigma_x + \Delta \tau_x\otimes\sigma_0\,,\\
T &=& -t\tau_z\otimes\sigma_0 + i\alpha\tau_z\otimes\sigma_z.
\label{eq2}
\end{eqnarray}
where $t$ is the hopping matrix element, $\alpha$ the spin orbit-coupling  and $B_x$ the Zeeman field. The Pauli matrices $\tau_{a}$ ($\sigma_{a}$) and the identity $\tau_0$ ($\sigma_0$) act in particle-hole (spin) space.

\begin{figure*}[t]
\includegraphics[width=0.7\textwidth]{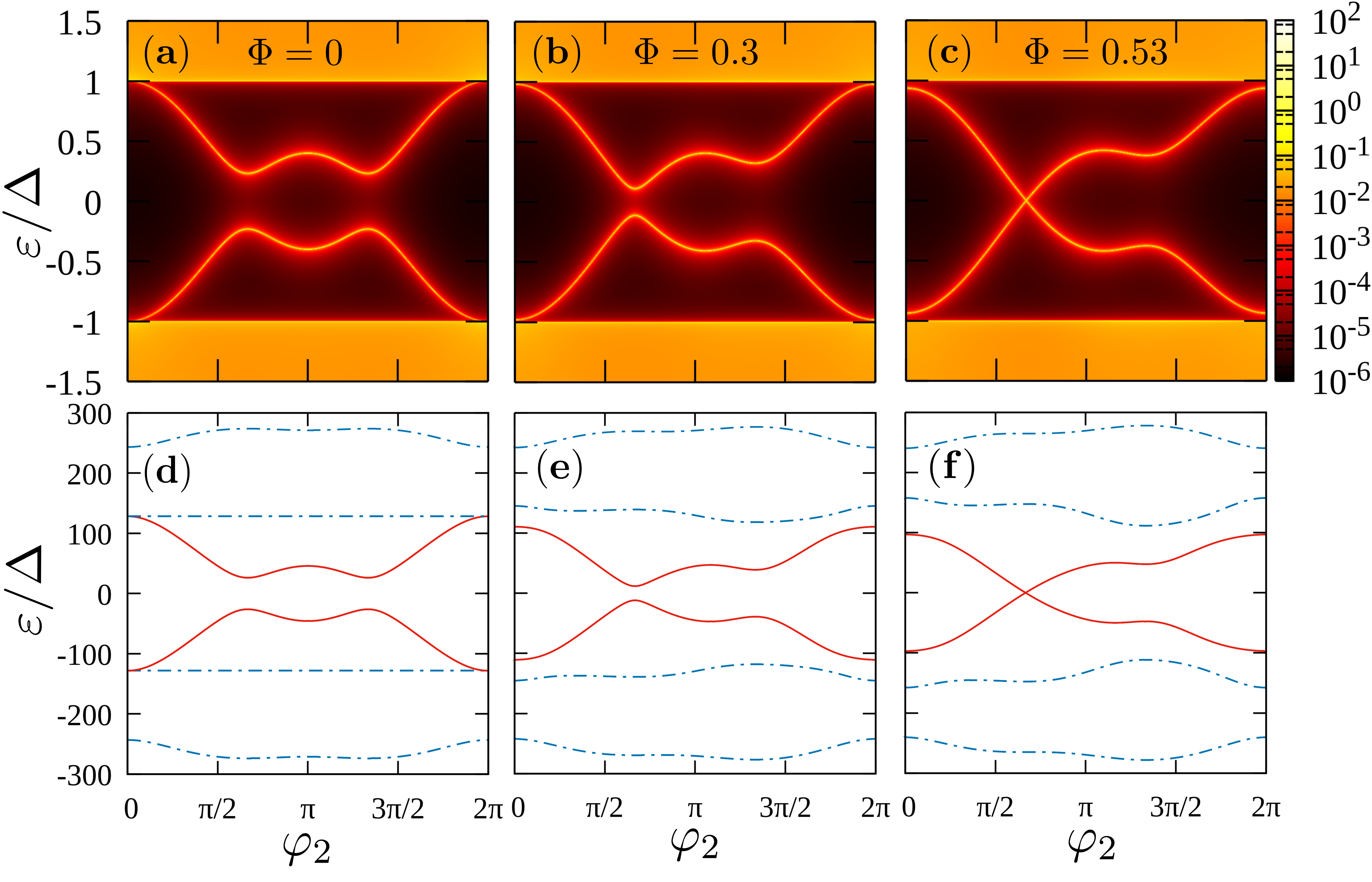}
\caption{Upper panels show the spectral density of the three terminal Josephson junction in the spin degenerate case ($B_x = 0$) along the path $\varphi_2 = -\varphi_1$ for different fluxes threading the central scattering region between the leads (a) $\Phi = 0$, (b) $\Phi=0.3$ and (c) $\Phi=0.53$. Lower panels (d), (e) and (f) show the eigenvalues of $H_{\text{top}}^{J}(\bm{\varphi})=-\mathcal{G}_{J}^{-1}(\omega=0)$ for this same path with the same corresponding fluxes. Solid lines indicate the bands topologically connected with the Andreev bound states and dashed lines indicate the eigenvalues bearing the topological information of the continuum states.}
\label{fig5}
\end{figure*}
The decimation procedure iteratively eliminates at each step of the Green's functions equations of motion of  even sites of the chain, eventually encapsulating numerous sites through a recursion relation using renormalized parameters. Assuming that the initial undressed Green's functions of the left boundary $g^{L}_{0}(\omega)$, the bulk $g^{B}_{0}(\omega)$, and the right boundary $g^{R}_{0}(\omega)$ are all the same and equal to $(\omega - \mathcal{H})^{-1}$, at step $n\geq1$ of the decimation they take the form
\begin{eqnarray}
g_n^{L}(\omega) &=& \left[(g_{n-1}^{L}(\omega))^{-1} - T_{n-1} g_{n-1}^{B}(\omega) T_{n-1}^{\dagger}\right]^{-1},\\
g_n^{B}(\omega) &=& \left[(g_{n-1}^{B}(\omega))^{-1} - \Sigma_{n-1}(\omega)\right]^{-1},\\
g_n^{R}(\omega) &=& \left[(g_{n-1}^{R}(\omega))^{-1} - T_{n-1}^{\dagger} g_{n-1}^{B}(\omega) T_{n-1}\right]^{-1},
\end{eqnarray}
where the renormalized hoppings are given by
\begin{eqnarray}
\notag
T_{n} &=& T_{n-1}g^{B}_{n-1}(\omega)T_{n-1},\\
T_{n}^{\dagger} &=& T_{n-1}^{\dagger}g^{B}_{n-1}(\omega)T_{n-1}^{\dagger}
\end{eqnarray}
and $\Sigma_{n-1}(\omega) =T_{n-1} g_{n-1}^{B}(\omega) T_{n-1}^{\dagger} + T_{n-1}^{\dagger} g_{n-1}^{B}(\omega) T_{n-1}$. The initial tunneling is just the undressed one $T_{n=0} = T$ [see Eq.~(\ref{eq2})]. For sufficiently large $n$, each of these Green's functions accurately represent the left and right boundary of a semi-infinite chain (generally not equal in topological systems) and the bulk of an infinite one. 

These boundary Green's functions are used to construct the three terminal junction Green's function (see Fig.~\ref{fig1}) as
$\mathcal{G}_J^{r/a}(\omega) = (g^{-1}(\omega\pm i\eta) - {\mathcal{T}})^{-1}$, where the supraindexes $r/a$ stand for retarded and advanced respectively, $g(\omega)$ is a $12\times 12$ block diagonal matrix of the form
\begin{equation}
g(\omega) = \left[ {\begin{array}{ccc}
   g^{R}(\omega) & \bm{0} & \bm{0}\\
   \bm{0} & g^{L}(\omega) & \bm{0} \\
   \bm{0} & \bm{0} & g^{R}(\omega)\\
\end{array} } \right],
\end{equation}
and $\mathcal{T}$ is the tunneling between the leads as described in Eq. (\ref{tunnelling}) in the main text. The Feynman Green's function used to calculate the winding number is then obtained as
\begin{equation}
\mathcal{G}_{J}(\omega) = \mathcal{G}^r_J(\omega) + f(\omega)\Big(\mathcal{G}_J^a(\omega)-\mathcal{G}_J^r(\omega)\Big),
\end{equation}
with $f(\omega)$ the Fermi-Dirac distribution function.
\section{Topological Hamiltonian for $B<B_c$: contributions from the ABS and the continuum states.}

It was recently stated that, under quite general assumptions, all the topological information of the total $\mathcal{G}(\omega)$ of a system is encoded in the the so called ``topological Hamiltonian" defined as $H_\mathrm{top}\equiv-\mathcal{G}^{-1}(\omega=0)$~\citep{Wang2012,Wang2013}. Notice this is a well defined hermitian operator, since the self energy $\Sigma(\omega=0)$ has no imaginary part. The key idea is that the entire $\mathcal{G}(\omega)$ can be smoothly deformed to $\mathcal{G}_{\text{eff}}(\omega) = [\omega + \mathcal{G}^{-1}(\omega=0)]^{-1}$ preserving its topology. In this case, it is easy to prove that the winding number of the Green's function  in a given parameter space $(\varphi_1,\varphi_2)$ can be calculated with the Chern numbers associated to each of the eigenstates $|\psi_{n}\rangle$ of $H_{\text{top}}$ as 
\begin{equation}
W_{12} = \sum\limits_{n\epsilon \text{occ}} \mathcal{C}^n_{12},
\end{equation} 
where the sum runs over all occupied states and
\begin{equation}
C_{12}^{n}\!=\!\frac{1}{2\pi}\!\int_{0}^{2\pi}\!d\varphi_1\!\int_{0}^{2\pi}\!d\varphi_2 \Big[i\langle\partial_{\varphi_1}\psi_{n}|\partial_{\varphi_2}\psi_{n}\rangle - (\varphi_1\! \leftrightarrow\!\varphi_2)\Big].
\end{equation}  
This theoretical tool provides a much simpler way of obtaining the topological invariants without the need of an integral over frequency space (see Eq. (\ref{W}) in the main text).

In the case of Josephson junctions, the eigenstates resulting from diagonalizing  $H_\mathrm{top}^{J}\equiv-\mathcal{G}_J^{-1}(\omega=0)$ should then reflect the topological nature of both the Andreev subgap structure and the continuum states above the superconducting gap, as we shall show below. There is only one caveat: the method is only applicable in our system for in plane magnetic fields below the critical value $B_x<B_c$, since for larger fields --after the topological transition-- an unpaired Majorana emerges at zero energy (the ``flat band"). In this case, the Green's function is singular at $\omega=0$ and hence, not invertible.

In the upper panels of Fig. \ref{fig5} we present the spectral density of the trijunction $\mathcal{A}(\omega)=-1/\pi\, \mathrm{Im}[\mathrm{Tr}(\mathcal{G}_J^r(\omega))]$ along the path $\varphi_2 = -\varphi_1$ for different values of the fluxes threading the central scattering region between the superconducting leads. The in plane magnetic field was chosen to be zero ($B_x=0$) so that the eigenenergies are spin degenerate. The lower panels show the outcome of diagonalizing the topological Hamiltonian for the corresponding fluxes. Six particle-hole symmetric and spin-degenerate bands are present. The eigenvalues in solid lines, which are closer to the Fermi level, should possess the topological information of the Andreev subgap states, while the dashed eigenvalues should represent the continuum states. Note that even though the energy scale is very different from the original one, the touching of the lowest ABS with the continuum at $\varphi_1 = \varphi_2 = 0$ and zero external flux $\Phi=0$ [see Figs. (a) and (d)] is accurately described, as well as the closing of the gap between the ABS at the first critical flux $\Phi = \Phi_c^{(1)} = 0.53$ [see Figs. (c) and (f)].

\begin{figure}[t]
\includegraphics[width=0.7\columnwidth]{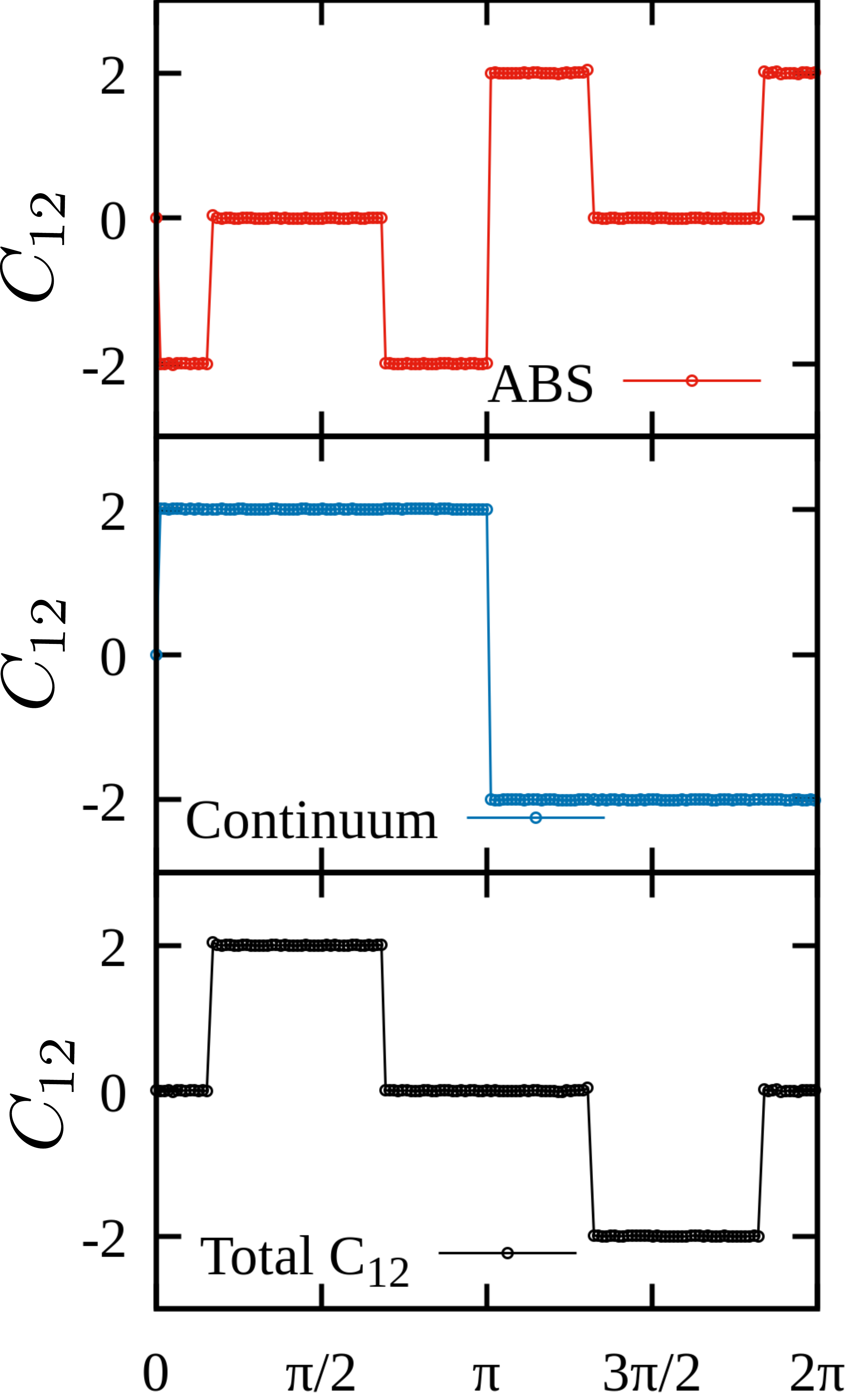}
\caption{Chern numbers $C_{12}$ of the Andreev bound states (ABS) and the continuum states. The lower panel shows the sum of both contributions.}
\label{fig6}
\end{figure}

In Fig.~\ref{fig6} we show the Chern numbers of these bands as a function of the external flux $\Phi$. 
The Chern associated to the ABS (upper panel in Fig.~\ref{fig6}) is obtained from the band closer to the Fermi level, while the one related to the continuum (middle panel in Fig.~\ref{fig6}) is just the sum of the Cherns of the two lower bands in Fig.~\ref{fig5}. All of them are spin degenerate and, as a result, they lead to a simple factor 2 in their respective topological invariants. The sum of this contributions (lower panel in Fig.~\ref{fig6}) is precisely the winding number of the full frequency-dependent Green's function obtained in the main text, revealing the validity of the method. Notice also that the contribution from the  ABS agree well with the calculation of Ref.~[\onlinecite{Meyer2017}].

This same procedure can be done for every $0<B_{x}<B_c$, where the spin degeneracy is lifted.  The topological Hamiltonian $H_{\text{top}}^J$ possesses in this case 12 different particle-hole symmetric eigenvalues. In this regime, the contribution from the continuum states and the Andreev levels cannot be separated, since the gap closing makes one of the ABSs merge with the continuum.  Nonetheless, the sum of all contributions up to the Fermi level correctly recovers the $s$-wave part of the phase diagram presented in Fig. 2(a) of the main text.

%
\end{document}